\documentclass[aps,prb,preprint,superscriptaddress,showpacs]{revtex4-1}
\usepackage{hyperref}
\usepackage{amsmath}
\usepackage{textcomp}
\usepackage{graphicx}

\begin{document}

\title{Non-Fermi-liquid behavior in cubic phase BaRuO$_{3}$: A dynamical mean-field study}
\author{Li Huang}
\affiliation{ Beijing National Laboratory for Condensed Matter Physics, 
              and Institute of Physics, 
              Chinese Academy of Sciences, 
              Beijing 100190, 
              China }
\affiliation{ Science and Technology on Surface Physics and Chemistry Laboratory, 
              P.O. Box 718-35, 
              Mianyang 621907, 
              Sichuan, 
              China }

\author{Bingyun Ao}
\affiliation{ Science and Technology on Surface Physics and Chemistry Laboratory, 
              P.O. Box 718-35, 
              Mianyang 621907, 
              Sichuan, 
              China }
\date{\today}

\begin{abstract}
Motivated by the recently synthesized cubic phase BaRuO$_{3}$ under high pressure 
and high temperature, a thorough study has been conducted on its temperature-dependent 
electronic properties by using the state-of-the-art \textit{ab inito} computing 
framework of density functional theory combined with dynamical mean-field theory. 
At ambient condition the cubic phase BaRuO$_{3}$ should be a weakly correlated 
Hund's metal with local magnetic moment. The spin-spin correlation function and 
local magnetic susceptibility can be well described by the Curie-Weiss law over 
a wide temperature range. The calculated low-frequency self-energy functions of 
Ru-$4d$ states apparently deviate from the behaviors predicted by Landau Fermi-liquid 
theory. Beyond that, the low-frequency optical conductivity can be fitted 
to a power-law $\Re\sigma(\omega) \sim \omega^{-0.98}$, which further confirms the 
Non-Fermi-liquid metallic state.
\end{abstract}


\maketitle

\section{introduction}

Recently the alkaline-earth ruthenium oxides have attracted growing interest. These
oxides generally exhibit fascinating physics properties, such as unconventional 
superconductivity with $p$-wave symmetry (Sr$_{2}$RuO$_{4}$),\cite{maeno:532} 
antiferromagnetic Mott insulator (Ca$_{2}$RuO$_{4}$),\cite{naka:2666} and 
orbital selective Mott transition in Sr$_{2-x}$Ca$_{x}$RuO$_{4}$,\cite{anisimov:2002} 
etc. Among the rest, due to their interesting magnetic phase diagrams, transport 
properties and potential device applications, the ternary ruthenates with perovskite 
or perovskite-related structures (ARuO$_{3}$: A = Ca, Sr, Ba) have been extensively 
studied by numerous experiments and theoretical calculations in the past 
decade.\cite{koster:253,cao:321,kos:2498,dodge:4932,lee:041104,laad:096402,
tak:060404,ahn:5321,maiti:161102,maiti:235110,jak:041103,had:203,lee:235113,cava:10005}

Both CaRuO$_{3}$ and SrRuO$_{3}$ crystallize in the orthorhombic perovskite structure 
with a GdFeO$_{3}$-type distortion. SrRuO$_{3}$ is a highly correlated, 
narrow-band metallic ferromagnet with a Curie temperature ($T_{c}$) of about 
160\ K.\cite{koster:253} Its local magnetic moment ($1.4\ \mu_{B}$) is rather large, 
despite highly extended $4d$ character of the valence electrons. Interestingly, 
CaRuO$_{3}$, an isostructural compound, does not show any magnetic ordering in 
finite temperatures.\cite{cao:321} The nature of its magnetic ground state still remains 
controversial. We note that one of the most striking properties of CaRuO$_{3}$ and 
SrRuO$_{3}$ compounds is the violation of Landau Fermi-liquid (LFL) theory, which 
has been proven by many experimental results, including x-ray photoemission spectra, 
transport and optical properties, etc.\cite{kos:2498,dodge:4932,lee:041104,laad:096402} 
The strength of Coulomb interaction $U$ and the importance of Hund's rule coupling $J$ 
among Ru-$4d$ orbitals are another two interesting topics and in lively debate. Though 
almost all the experimental\cite{tak:060404,ahn:5321,maiti:161102} and theoretical\cite{had:203,
jak:041103,maiti:235110} efforts manifest some role of electron-electron correlation, 
the strength and the extent of its importance still remain unclear.

In the earlier years, it is well known that depending on how BaRuO$_{3}$ is synthesized
it has several polytype structures, i.e., the nine-layered rhombohedral (9R), the four 
layered hexagonal (4H), and the six layered hexagonal (6H).\cite{cava:10005,lee:235113} 
Lately, the cubic phase BaRuO$_{3}$ with ideal perovskite structure has been synthesized by 
Jin \emph{et al.} under 18\ GPa at 1000 \textcelsius.\cite{jin:2008} It remains metallic 
down to 4\ K and occurs a ferromagnetic transition at $T_{c} = 60$\ K,\cite{zhou:077206} 
which is significantly lower than that of SrRuO$_{3}$.\cite{koster:253} The ferromagnetic 
transition in SrRuO$_{3}$ falls into the mean-field universality class whereas cubic 
phase BaRuO$_{3}$ exhibits significant critical fluctuations as described by the 3D 
Heisenberg model.\cite{zhou:077206} The availability of cubic phase BaRuO$_{3}$ not 
only completes the polymorph of BaRuO$_{3}$, but also makes it possible to map out the 
evolution of magnetism and other properties as a function of the ionic size of the 
A-size in the whole series of ARuO$_{3}$.\cite{jin:2008}

Despite tremendous efforts have been made, little is known about the basic properties 
of cubic phase BaRuO$_{3}$. In this paper, we will address the following two issues:
(i) Definitely, in CaRuO$_{3}$ and SrRuO$_{3}$, the effects induced by electronic 
correlation can not be ignored.\cite{had:203,maiti:161102,tak:060404,jak:041103,
ahn:5321,maiti:235110} However, does the electronic correlation play a decisive role in 
cubic phase BaRuO$_{3}$? (ii) It is believed that the physical properties
of CaRuO$_{3}$ and SrRuO$_{3}$ can not be well described by the LFL theory. Indeed, 
the evidences are their low-frequency optic conductivity, resistivity, and electronic 
Raman scattering intensity which obey the fractional power-law.\cite{lee:041104,
kos:2498,dodge:4932,laad:096402} Thus, whether the physical properties of cubic 
phase BaRuO$_{3}$ still violate the LFL theory is an essential problem.

The density functional theory within local density approximation combined with 
dynamical mean-field theory (dubbed LDA + DMFT) is a very powerful computing framework 
for strongly correlated materials.\cite{antoine:13,kotliar:865,amadon:205112} In the 
present works, by employing the LDA + DMFT computational scheme, the temperature-dependent 
electronic properties of cubic phase BaRuO$_{3}$ have been systematically studied. In 
contrast to CaRuO$_{3}$ and SrRuO$_{3}$, under room temperature cubic phase BaRuO$_{3}$ 
is a weakly correlated Hund's metal with local magnetic moment, and its low-frequency 
conductivity deviates the $\omega^{-2}$ law as is predicted by classic LFL theory.

\section{method}
We first compute the ground state electronic structures of cubic phase BaRuO$_{3}$ 
within nonmagnetic configuration by utilizing the plane-wave pseudopotential approach 
as is implemented in the {\scriptsize QUANTUM ESPRESSO} software package.\cite{qe2009} 
The generalized gradient approximation with Perdew-Burke-Ernzerhof 
exchange-correlation functional\cite{pbe:3865} is used to describe the exchange 
and correlation potentials. The pseudopotentials in projector augmented wave 
scheme\cite{paw:17953} for Ba, Ru, and O species are built by ourselves. The 
electronic wave functions are described with a plane-wave basis truncated at 80 Ha, 
and a $\Gamma$-centered $12 \times 12 \times 12$ $k$-point grid for Brillouin zone 
integrations is adopted. These pseudopotentials and computational parameters are 
carefully checked and tuned to ensure the numerical convergences.

To include the effect of electronic correlation, the ground state wave functions 
are used to construct a basis of maximally localized Wannier functions (MLWF) for 
Ru-$4d$ and O-$2p$ orbitals. The corresponding multiband Hubbard Hamiltonian has 
the following form\cite{kotliar:865,amadon:205112}
\begin{equation}
\label{eq:lda_dmft_ham}
H_{\text{LDA+DMFT}} = H_{\text{LDA}} - H_{\text{DC}} 
+\sum_{imm^{\prime}} \frac{U_{mm^{\prime}}}{2} n_{im}n_{im^{\prime}},
\end{equation}
where $n_{im} = c^{\dagger}_{im}c_{im}$, and $c^{\dagger}_{im}(c_{im})$ creates 
(destroys) an electron in a Wannier orbital $m$ at site $i$. Here $H_{\text{LDA}}$ 
is the effective low-energy Hamiltonian in the basis of Ru-$4d$ and O-$2p$ MLWF 
orbitals, and thus is a $ 14 \times 14 $ matrix. $H_{\text{DC}}$ is a double counting 
correction term which accounts for the electronic correlation already described 
by the LDA part, and the fully local limit scheme\cite{amadon:155104} is chosen. 
The Coulomb interaction is taken into considerations merely among the Ru-$4d$ 
orbitals. We use $U = 4.0$\ eV and $J = 0.65$\ eV to parameterize the Coulomb 
interaction matrix, which are close to previous estimations.\cite{jak:041103,had:203} 
To solve the many-body Hamiltonian (\ref{eq:lda_dmft_ham}), in the DMFT part\cite{kotliar:865,antoine:13} 
we employ the hybridization expansion continuous time quantum Monte Carlo impurity 
solver (abbreviated CT-HYB).\cite{werner:076405,gull:349} Finally, through the mature 
analytical continuation methods\cite{jarrell:133,beach} the impurity spectral function 
can be extracted directly from the imaginary-time Green's function which are derived 
from the quantum Monte Carlo simulations.

\section{results and discussion}

\begin{figure}
\centering
\includegraphics[scale=0.65]{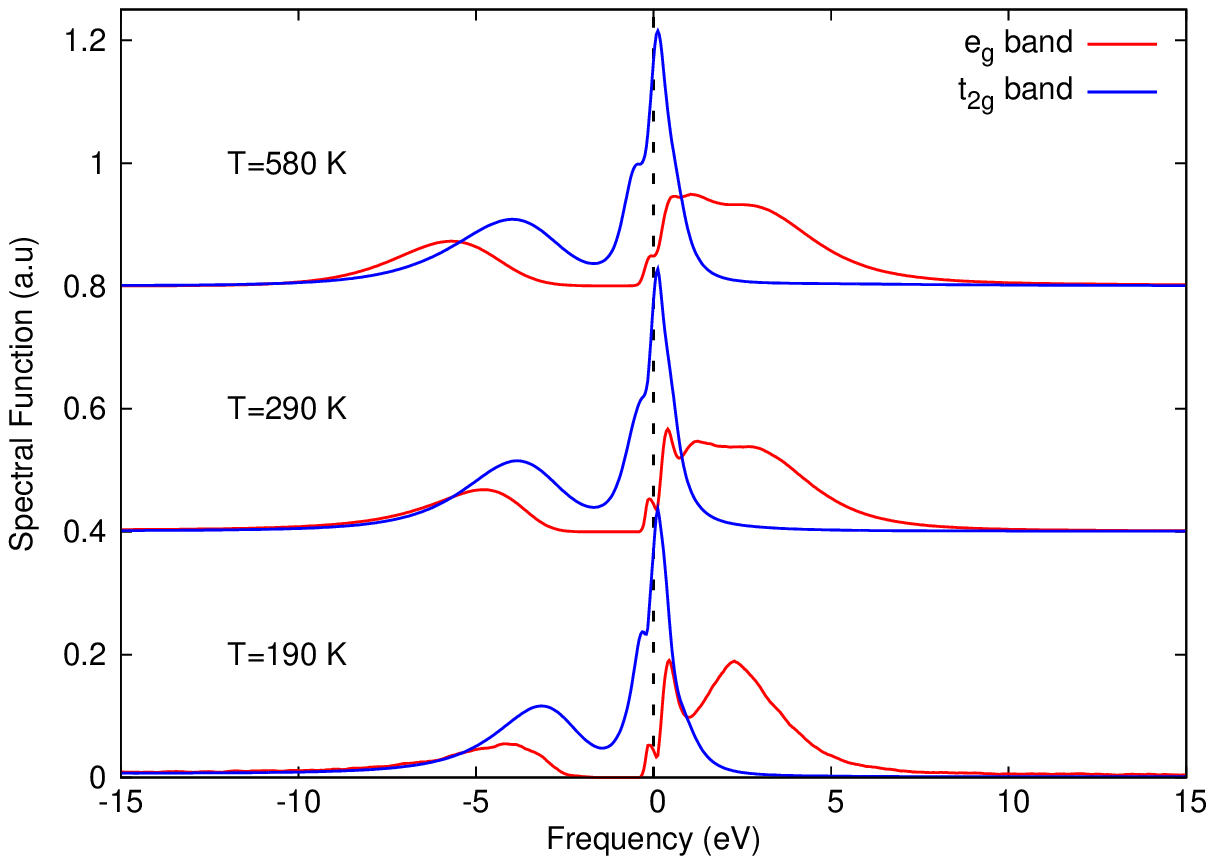}
\includegraphics[scale=0.75]{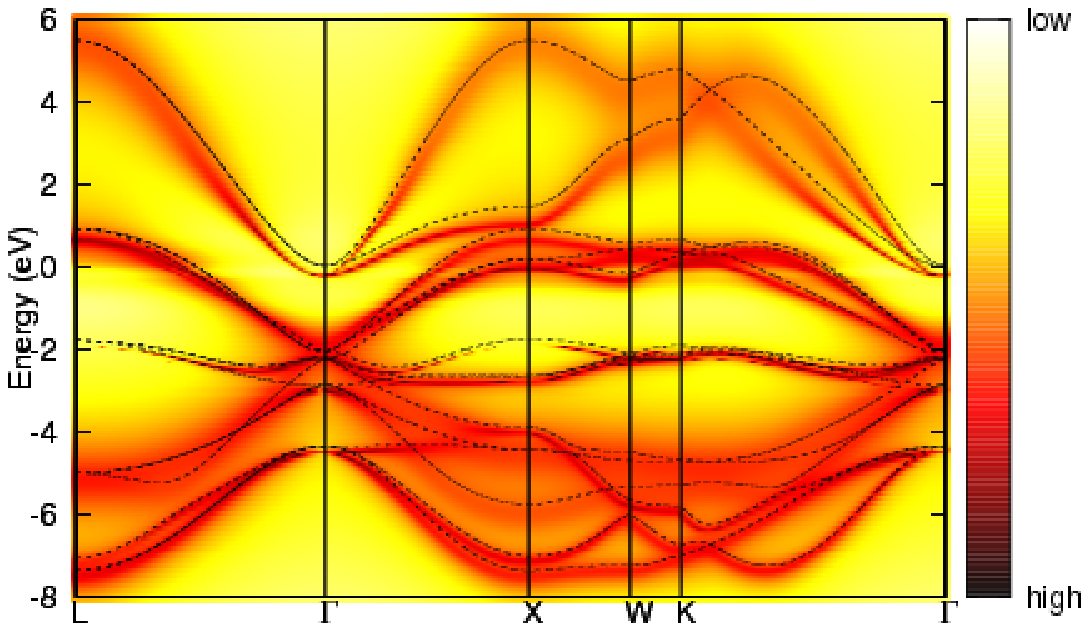}
\caption{(Color online) Spectral functions of cubic perovskite BaRuO$_{3}$ calculated 
by LDA + DMFT method. Upper panel: Spectral functions of Ru-$4d$ states at various 
temperatures. Lower panel: Quasiparticle band structure of BaRuO$_{3}$ along 
high symmetry lines in the Brillouin zone.\label{fig:quasi_band}}
\end{figure}

Figure \ref{fig:quasi_band} represents our calculated orbital-resolved density of 
states for Ru-$4d$ states at several typical temperatures. The octahedral surrounding 
of Ru splits the Ru-$4d$ states into three-fold degenerated $t_{2g}$ and two-fold 
degenerated $e_{g}$ levels. Ru$^{4+}$ ion, 
which is nominally in a low-spin, $d^{4}$ configuration, gives rise to a $t^{4}_{2g}$ 
configuration with Fermi level lying in the $t_{2g}$ manifold with empty $e_{g}$ states.
As for the density of states of $t_{2g}$ states, it displays a sharp quasiparticle 
peak near the Fermi level, a shoulder structure around -0.3\ eV, and a Hubbard 
subband like hump at -8.0\ eV $\sim$ -2.0\ eV. While for the density of states of $e_{g}$ 
states, since it is less occupied, the primary spectral weight is above the Fermi 
level. There are two small satellites located on both sides of the Fermi level 
(-0.1 and 0.3\ eV, respectively). With the increment of temperature, the two peaks will be 
smeared out gradually. To sum up, the integrated spectral functions of Ru-$4d$ 
states show significantly metallic features, and the temperature effect is not 
very obvious. When the temperature rises from 190\ K to 580\ K, slightly spectral 
weight transfer to high energy is observed.

In the next step, we computed the full momentum-resolved spectral function 
$A(k,\omega)$ along some high symmetry lines in the Brillouin zone for cubic phase 
BaRuO$_{3}$. The inverse temperature $\beta$ is chosen to be 40, which
corresponding to $T = 290$\ K approximately. In the lower panel of Fig.\ref{fig:quasi_band} 
$A(k,\omega)$ is shown in comparison with the LDA band structure. The sharp 
quasiparticle peak observed in the integrated spectral function is clearly 
visible on the intensity plot, and fairly well defined. It lies in the region 
about from -2.0\ eV to 1.5\ eV, dominated by the $t_{2g}$ states. At higher 
energy, the $e_{g}$ states become the majority. However, in the region below 
-2.0\ eV, the O-$2p$ states make a major contribution. From the distribution 
of spectral weights of $t_{2g}$ and $e_{g}$ states, it is speculated that in 
the region from -2.0\ eV to -7.0\ eV, there exists strong hybridization between 
the Ru-$4d$ and O-$2p$ states. Comparing this with the LDA band structure, first 
of all we notice the quasiparticle band structure does not show apparent shifting. 
Secondly, except for becoming diffuse, the band renormalization of the quasiparticle 
band structure is hard to be distinguished. In general, the quasiparticle band 
structure of cubic phase BaRuO$_{3}$ coincides with its LDA band structure, giving 
rise to a picture of weakly correlated metal. On the contrary, previous LDA + DMFT 
calculations for CaRuO$_{3}$ and SrRuO$_{3}$ present strongly renormalized and 
shifted quasiparticle band structure,\cite{jak:041103} resulting in the picture 
of moderately correlated metal.


\begin{figure}
\centering
\includegraphics[scale=0.65]{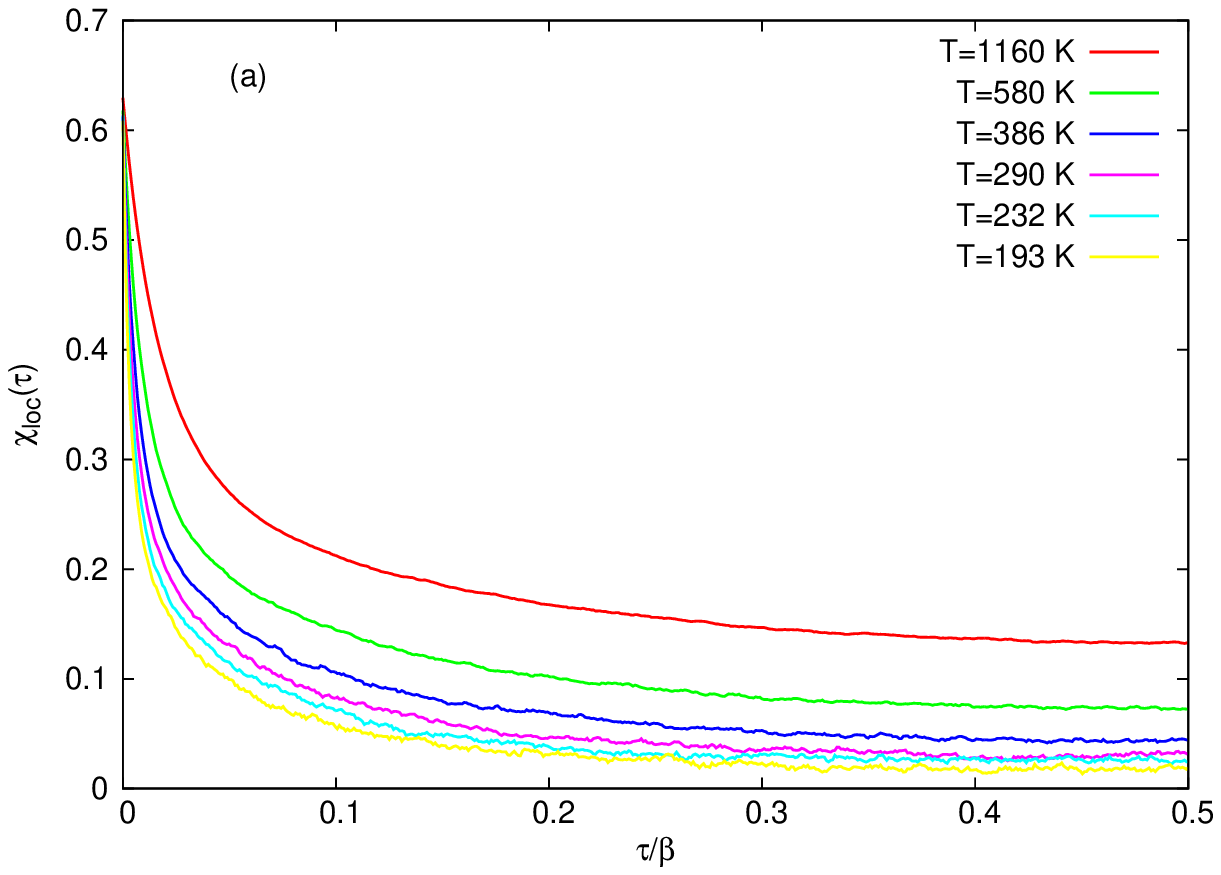}
\includegraphics[scale=0.65]{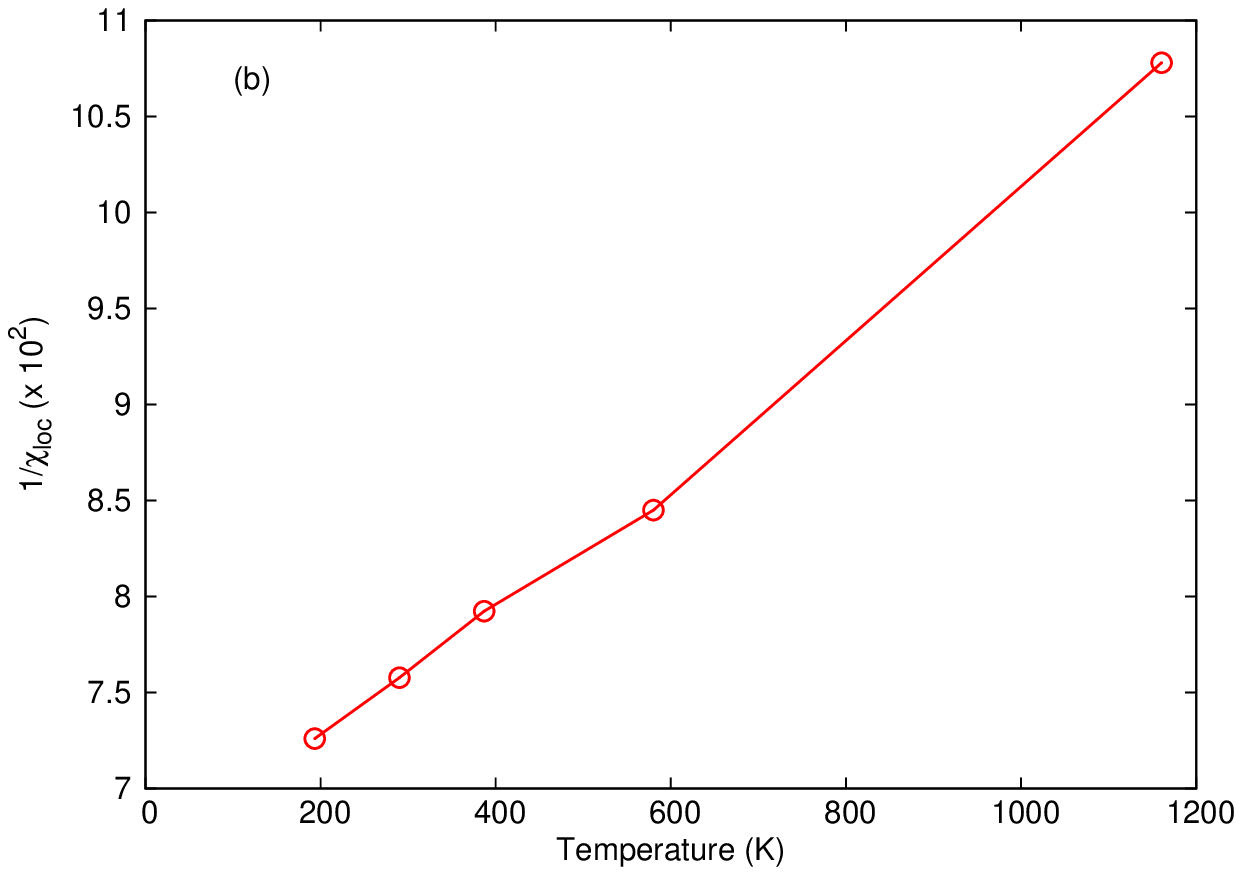}
\caption{(Color online) Magnetic properties of cubic perovskite BaRuO$_{3}$. (a) 
Spin-spin correlation functions $\chi(\tau) = \langle S_{z}(0) S_{z}(\tau) 
\rangle$ at various temperatures. (b) Inverse local magnetic susceptibility 
$\chi^{-1}_{loc}$ as a function of temperature. \label{fig:schi}}
\end{figure}

In recent years, the evolutional trend of ferromagnetism in ARuO$_{3}$ is in hot 
debate.\cite{cao:321,jin:2008,zhou:077206} Thus in the present works, we calculated 
the spin-spin correlation function $\chi(\tau)$ and local magnetic susceptibility 
$\chi_{loc}$ of cubic phase BaRuO$_{3}$, and tried to elucidate its magnetic properties 
in finite temperatures. The calculated spin-spin correlation functions are illustrated 
in Fig.\ref{fig:schi}(a). On one hand, the cubic phase BaRuO$_{3}$ exhibits a well-defined 
frozen local moment, which is characterized by a spin-spin correlation function that 
approaches non-zero constants at large enough $\tau$, as is easily seen from $T = 
193$\ K to 1160\ K. On the other hand, the spin-spin correlation function does not 
behave as $\chi(\tau) \sim (T/\sin(T\tau\pi))^2$ for times $\tau$ sufficiently 
far from $\tau = 0$ or $\beta$ respectively, which means the violation of 
LFL theory.\cite{werner:166405} From the spin-spin correlation function, the local 
magnetic susceptibility $\chi_{loc} = \int^{\beta}_{0} \chi(\tau) \text{d}\tau $ can 
be easily evaluated, which is plotted in Fig.\ref{fig:schi}(b). As shown, the calculated 
$\chi_{loc}$ is Curie-Weiss like over a rather wide temperature range, in other words, 
it follows a $\chi^{-1}_{loc}(T) = T/C$ law at least up to $T = 1160$\ K. This implies 
that the Ru-$4d$ electrons in cubic phase BaRuO$_{3}$ retain the local nature of the magnetic 
moment.


\begin{figure}
\centering
\includegraphics[scale=0.65]{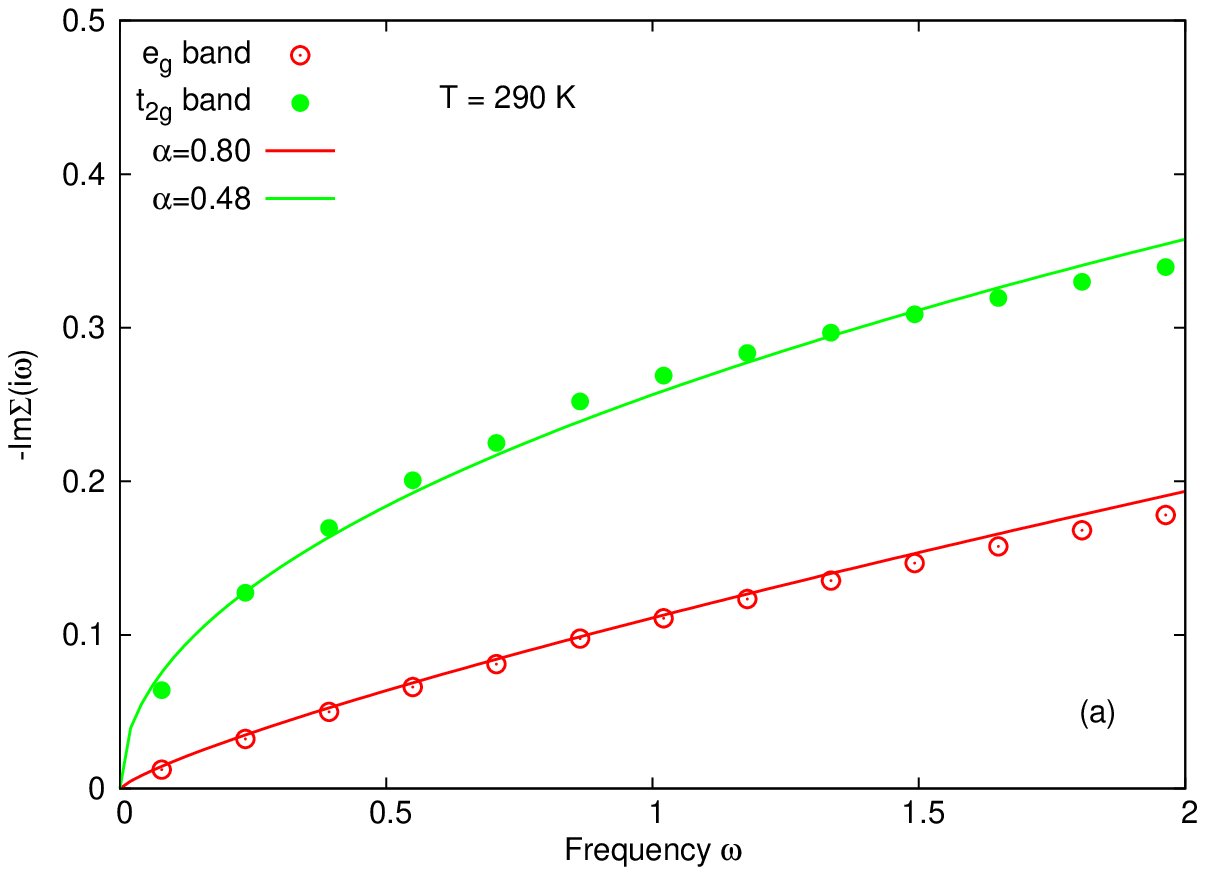}
\includegraphics[scale=0.65]{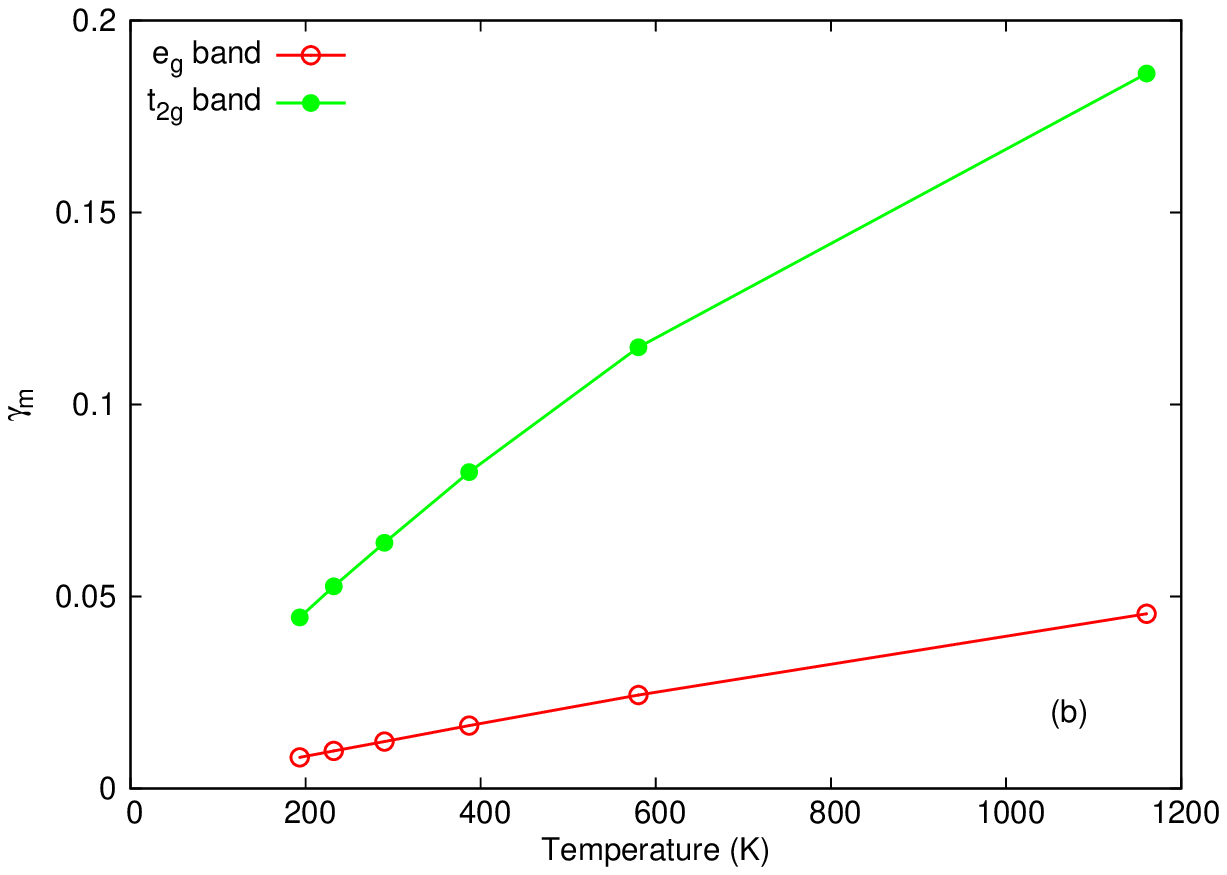}
\caption{(Color online) Electronic self-energy function of Ru-$4d$ states. (a) Imaginary 
part of the Matsubara self-energy function $\Im\Sigma(i\omega)$ for $t_{2g}$ and $e_{g}$ 
orbitals at $T = 290$\ K. The solid lines denote as the fitted function $-\Im\Sigma(i\omega) = 
A(i\omega)^{\alpha} + B$. (b) Orbital-resolved low-energy scattering rate 
$\gamma_{m} = -\Im\Sigma_{m}(i\omega \rightarrow 0)$. \label{fig:nfl}}
\end{figure}

Next we concentrate our attentions to the electronic self-energy functions of Ru-$4d$ 
states. The calculated orbital-resolved $\Im\Sigma(i\omega)$ are shown in Fig.\ref{fig:nfl}(a). 
For the sake of simplicity, only those results calculated at $T = 290$\ K are presented. 
Werner \emph{et al.}\cite{werner:166405} have suggested that the still-mysterious optical 
conductivity $\sigma(\omega)$ in pseudocubic SrRuO$_{3}$ and CaRuO$_{3}$, which varies
approximately as $\omega^{-0.5}$ and deviates sharply from the prediction of LFL theory,
can be perfectly interpreted as a consequence of square-root self-energy function. Inspired by 
their works, we conducted a careful analysis to determine the asymptotically formula for 
the low-frequency self-energy function. In a Fermi-liquid, the imaginary part of Matsubara 
self-energy should exhibit a linear regime at low energy, whose slope is directly related 
to the quasiparticle mass enhancement. However, as shown in Fig.\ref{fig:nfl}(a), we do 
not observe any linear behavior: the Matsubara self-energy behaves as $-\Im\Sigma(i\omega) 
= A(i\omega)^{\alpha} + \gamma$ with $\alpha \sim 0.48 $ for $t_{2g}$ states and $\alpha 
\sim 0.80$ for $e_{g}$ states, respectively. The non-linear frequency dependence of
the Matsubara self-energy implies that Landau quasiparticles and effective masses can not 
be properly defined for cubic phase BaRuO$_{3}$. The non-zero intercept $\gamma = -\Im\Sigma(i\omega 
\rightarrow 0)$ can be viewed as the low-energy scattering rate and it is a broadly used 
physical quantity to distinguish the LFL and NFL phases.\cite{werner:166405} As a byproduct, 
the orbital-resolved $\gamma_{m}$ is evaluated as a function of temperature and shown in 
Fig.\ref{fig:nfl}(b). Clearly, $\gamma_{m}$ increases monotonously with the increment of 
temperature and $\gamma_{t_{2g}} > \gamma_{e_{g}}$ is always valid. For both $t_{2g}$ and 
$e_{g}$ states $\gamma_{m}$ can not be neglected even at $T = 190$\ K. Thus, it means 
that similar to SrRuO$_{3}$ and CaRuO$_{3}$, the cubic phase BaRuO$_{3}$ lies in the NFL 
regime as well.


\begin{figure}
\centering
\includegraphics[scale=0.65]{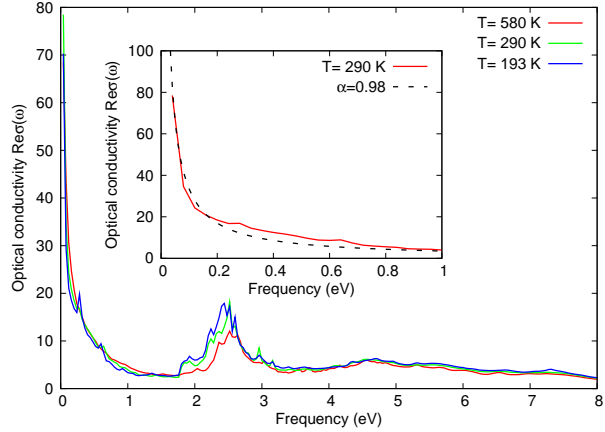}
\caption{(Color online) Real part of optical conductivity of cubic perovskite BaRuO$_{3}$ by 
LDA + DMFT calculations. Inset: The low-frequency $\Re\sigma(\omega)$ at $T = 290$\ K, and the 
dashed line represents the fitted function $\Re\sigma(\omega) = C\omega^{-\alpha}$. \label{fig:optic}}
\end{figure}

A power-law analysis on the transport properties, such as resistivity $\rho(T) \propto 
T^{n}$, of cubic phase BaRuO$_{3}$ was made by Zhou \emph{et al}.\cite{zhou:077206} and 
the exponent $n$ as a function of pressure was evaluated recently. Their results show an interesting
evolution from $n \sim 1.85$, which is close to $n = 2$ for the LFL phase at ambient 
pressure, to $n \sim 1.4$ of the NFL phase at the pressure where the ferromagnetic phase 
collapses. The most important evidence for NFL state in SrRuO$_{3}$ and CaRuO$_{3}$ is
the fractional power-law conductivity.\cite{kos:2498,dodge:4932,lee:041104} In this work, 
we also calculate the optical conductivity $\sigma(\omega)$ of cubic phase BaRuO$_{3}$ 
under various temperatures. In Fig.\ref{fig:optic} only the real part of optical conductivity 
is shown. The sharp peak near $\omega = 0$ denotes the Drude-like feature. The broad hump
located from 1.5\ eV to 3.5\ eV can be attributed to the contribution of electron transition 
between quasiparticle peak and Hubbard subbands.\cite{ahn:5321} With the increment of 
temperature, this hump slightly shifts to higher frequency region, which is in accord 
with the variation trend of Hubbard subbands observed in the temperature-dependent 
integrated spectral functions of Ru-$4d$ states (see Fig.\ref{fig:quasi_band}). In order 
to further confirm whether the underlying physics of cubic phase BaRuO$_{3}$ can be described with 
LFL theory, we conduct a detailed power-law analysis for the low-frequency optical 
conductivity at $T = 290$\ K. The low-frequency optical conductivity is fitted by the 
exponent function $\Re\sigma(\omega) \sim C\omega^{-\alpha}$. The quantitative results are 
shown in the inset of Fig.\ref{fig:optic}. The fitted exponent $\alpha \sim 0.98$, while 
the expected value predicted by LFL theory is $\alpha = 2$. It's worth mentioning that 
the exponent $\alpha$ is approximately 0.5 for pseudocubic SrRuO$_{3}$ and CaRuO$_{3}$, 
and 0.7 for some high temperature superconductivity cuprates.\cite{kos:2498,dodge:4932}
Nevertheless, the optical conductivity data suggest the NFL metallic nature of cubic phase 
BaRuO$_{3}$ under ambient condition again.

Finally, we should emphasize the importance of Hund's physics in cubic phase BaRuO$_{3}$.
Very recent investigations about iron pnictides and chalcogenides showed that strong 
correlation is not always caused by the Hubbard interaction $U$, but can arise from 
the Hund's rule coupling $J$.\cite{haule:025021,yin:2011} Since the strength of electronic 
correlation in these materials is almost entirely due to the Hund's rule coupling, they 
are dubbed Hund's metals by Haule \emph{et al.}\cite{haule:025021} at first. It has 
recently been noticed by Yin \emph{et al.} that in realistic Hund's metals, the 
electronic self-energy and corresponding optical conductivity show NFL power-law 
frequency dependence, tendency towards strong orbital differentiation, and that large 
mass enhancement can occur even though no clear Hubbard subband exist in the single 
particle spectra.\cite{werner:166405,yin:2011,yin:2012,kute:045105} According to their 
investigations, both the iron pnictides and chalcogenides are typical Hund's metals. 
The origin of fractional power-law in the optical conductivity of them can be traced 
to the Hund's rule interaction. As for the cubic phase BaRuO$_{3}$, based on our 
calculated results: NFL behavior in low-frequency self-energy function and scattering 
rate, fractional power-law in the optical conductivity, and considerable mass 
enhancement (at $T = 290$\ K $m^{*}_{t_{2g}} = 1.8m_{0}$ and $m^{*}_{e_{g}} = 1.2m_{0}$), we 
can conclude that it is another realistic Hund's metal. Indeed, we have performed additional 
LDA + DMFT calculations for cubic phase BaRuO$_{3}$ with different Coulomb interaction 
strengths from $U$ = 2.0\ eV to 6.0\ eV and obtained almost identical results. However, 
when $U = 4.0$\ eV and the Hund's rule coupling term is completely ignored ($J = 0.0$\ eV), 
the NFL behaviors previously found in the self-energy function and optical conductivity
are absent totally. It should be noted that Medici \emph{et al.}\cite{medici:256401} 
have suggested that the physical properties of ruthenates are governed by the Hund's 
physics, in other words, the pervoskite ARuO$_{3}$ forms a new series of Hund's metal. Our 
calculated results for the cubic phase BaRuO$_{3}$ confirm their issue as well.

\section{summary}
In summary, to find out a consistent description for the ARuO$_{3}$-type ruthenates,
we study the temperature-dependent physical properties of recently synthesized cubic
phase BaRuO$_{3}$ by using the first principles LDA + DMFT approach. Judged from the 
quasiparticle band structure and integrated spectral functions of Ru-$4d$ states, 
the cubic phase BaRuO$_{3}$ is a weakly correlated Hund's metal. There exists local 
magnetic moment and the inverse local magnetic susceptibility obeys the Curie-Weiss 
law in the studied temperature regime. The low-frequency self-energy function, 
scattering rate, and optical conductivity of cubic phase BaRuO$_{3}$ show apparent 
NFL behaviors. It is argued that the Hund's rule coupling $J$ plays an important 
role in the underlying physics in cubic phase BaRuO$_{3}$ and other perovskite 
ARuO$_{3}$ compounds.

\begin{acknowledgments}
LH was supported by the National Science Foundation ͑of China and that
from the 973 program of China under Contract No.2007CB925000 and No.2011CBA00108. BYA was
supported by the National Science Foundation ͑of China under Contract No.20971114.
\end{acknowledgments}

\bibliography{bro}

\end{document}